\documentclass[useAMS,usenatbib]{mn2e}
\usepackage{graphicx}
\usepackage[figuresright]{rotating}
\usepackage{subfigure}
\usepackage{longtable}
\usepackage{txfonts}

\newcommand\ion[2]{#1$\,${\scshape{#2}}}

\title[Analyzing derived metallicities and ionization parameters from
model-based determinations]{Analyzing derived metallicities and ionization parameters from
model-based determinations in ionized gaseous nebulae}

\author[Dors  et al.]
{O.~L. Dors Jr.$^{1}$\thanks{E-mail:olidors@univap.br}, Angela Krabbe$^1$, Guillermo F. H\"agele$^{2,3}$, 
Enrique P\'erez-Montero$^{4}$ \\
$^1$ Universidade do Vale do Para\'iba, Av. Shishima Hifumi, 2911, Cep
12244-000,
S\~ao Jos\'e dos Campos, SP, Brazil\\
$^2$Facultad de Ciencias Astron\'omicas y Geof\'{\i}sicas, Universidad Nacional de la La Plata, Paseo del Bosque s/n, 1900 La Plata, Argentina.\\
$^{3}$ Departamento de F\'{\i}sica Te\'orica, C-XI, Universidad Aut\'onoma de Madrid, 28049 Madrid, Spain.\\
$^{4}$ Instituto de Astrof\'sica de Andaluc\'ua (CSIC), PO Box 3004, 18080 Granada, Spain}

\begin{document}

\date{Accepted -. Received -.}

\pagerange{\pageref{firstpage}--\pageref{lastpage}} \pubyear{2011}

\maketitle

\label{firstpage}

\begin{abstract}
  We analyze the reliability  of oxygen abundances and ionization parameters obtained
from different diagnostic diagrams. For this, we compiled from the literature observational emission line intensities 
and oxygen abundance of  446  star-forming regions whose O/H abundance  was determined by direct estimation of electron temperature. The  abundances compiled  were compared with the values calculated in this work using  different  diagnostic diagrams in combination  
with results from a grid of photoionization models. 
We found that  the [\ion{O}{iii}]/[\ion{O}{ii}] vs. [\ion{N}{ii}]/[\ion{O}{ii}], 
[\ion{O}{iii}]/H$\beta$ vs. [\ion{N}{ii}]/[\ion{O}{ii}], and ([\ion{O}{iii}]/H$\beta$)/([\ion{N}{ii}]/H$\alpha$) vs. 
[\ion{S}{ii}]/[\ion{S}{iii}] diagnostic diagrams   
give O/H values close to the $T_{\rm e}$-method, with  differences    
of about 0.04 dex and dispersion of about 0.3 dex. Similar results were obtained by detailed models 
but with a   dispersion of 0.08 dex. 
The origin of the dispersion found in the use of diagnostic diagrams  is    probably due
to differences between the real N/O-O/H relation of the sample and the one assumed in the
models. This is confirmed by the use of detailed models that do not have a fixed
N/O-O/H relation. We found no correlation  between ionization parameter and the metallicity for the 
objects of our sample.
We conclude that  the combination
of two line ratio predicted by photoionization models, one sensitive to the metallicity and another
sensitive to the ionization parameter, which  takes into account 
the physical conditions  of  star-forming regions,  
gives O/H estimates close to the values derived using direct detections of electron temperatures.
\end{abstract}

\begin{keywords}
galaxies: general -- galaxies: evolution -- galaxies: abundances --
galaxies: formation-- galaxies: ISM
\end{keywords}


\section{Introduction}


Oxygen  abundance estimates in star-forming regions play a 
crucial role in the understanding of galaxy evolution.
For example, oxygen radial gradients in spiral galaxies obtained
by \ion{H}{ii} region observations
(e.g. \citealt{stanghellini10};  \citealt{kewley10}; \citealt{bresolin09}; \citealt{krabbe08};
\citealt{dors05}; \citealt{kennicutt03}) 
are essential to test chemical
evolution models (see \citealt{molla05}) and to 
investigate the effect of environment on  galaxy interactions \citep{ellison10,dors06,skillman96} 
as well as the mass-metallicity relation of galaxies (e.g. \citealt{pyliugin04}; \citealt{perez09}).
Likewise, oxygen abundance estimates in metal-poor galaxies are also important to test theories of chemical evolution
of galaxies because these are the least chemically evolved objects \citep{kunt83}.

Unfortunately, for the most of   star-forming regions, only collisionally excited
emission-lines (CELs) in the optical are bright enough to be used for the derivation
of elemental abundance.  CELs are temperature sensitive, thus, only an accurate determination
of the metallicity can be achieved from the previous estimation of the electron temperature 
(this method will be called $T_{\rm e}$-method) using, for instance, the ratio of different
CELs [\ion{O}{iii}]$(\lambda4959+\lambda5007)/\lambda4363$ which are weak or unobservable
in star-forming regions with high metallicity and/or low excitation 
\citep{dors08,diaz07}. In these cases,  oxygen abundances can be obtained
by empirical (i.e. using  oxygen determinations via $T_{\rm e}$-method) or theoretical
(i.e. using photoionization models) calibrations   between oxygen abundances and more easily measured
line ratios (hereafter  strong-line methods). The oxygen abundance indicator
$R_{23}$=([\ion{O}{ii}]$\lambda$3727+[\ion{O}{iii}]$\lambda$4959,$\lambda$5007)/H$\beta$
proposed by \citet{pagel79} has found large acceptance in this context
and several authors have calibrated this line ratio with O/H abundance
(e.g. \citealt{edmunds84}; \citealt{dopita86}; \citealt{pilyugin01}; \citealt{dors05}; among others). 
Additional O/H indicators based on other emission lines
such as $N_{2}$=[\ion{N}{ii}]$\lambda$6584/H$\alpha$ \citep{thaisa94}, 
[\ion{N}{ii}]$\lambda$6584/[\ion{O}{iii}]$\lambda$5007  \citep{alloin79},
$S_{23}$=([\ion{S}{ii}]$\lambda\lambda$6716,6731+[\ion{S}{ii}]$\lambda\lambda$9069,9532)/H$\beta$
\citep{vilchez96, diaz00} have also been suggested (see also \citealt{kewley02}). 
However, distinct methods  
or distinct calibrations of a same oxygen indicator provide different oxygen values with 
discrepancies up to 1.0 dex  \citep{kewley08,rupke08,dors05,kennicutt03}. Currently, the large
number of direct oxygen estimates available in the literature have helped
to investigate this discrepancy. For example, \citet{yin07}
determined the gas-phase oxygen abundance for a sample of 695 galaxies and
\ion{H}{ii} regions using the $T_{\rm e}$-method and compared these determinations
with the ones via $R_{23}$,   $N_{2}$, 
([\ion{N}{ii}]$\lambda$6584/H$\alpha$)/([\ion{O}{iii}]$\lambda$5007/H$\beta$),
and [\ion{S}{ii}]($\lambda$6717+$\lambda$6731)/H$\alpha$. They found that
among the indices above, the  $N_{2}$ provides
more consistent O/H abundances when compared with the ones via $T_{\rm e}$-method. 
Similar analysis was also done by \citet{perez05},
\citet{liang06}, and  \citet{nagao06}.

The studies above analyzed  strong-line methods  based mainly on one
line ratio, such as the $R_{23}$,  $N_{2}$, among others.
In principle, the use of diagnostic diagrams, suggested by 
\citet{baldwin81} to separate objects according to their primary excitation
mechanisms, containing line ratios
strongly dependent on the degree of ionization and on the metallicity of  star forming regions,
can  improve the accuracy of strong-line methods.  Although  
a large number of these diagrams have been applied to estimate
oxygen abundances and ionization parameters of star forming regions (e.g. \citealt{levesque10}, \citealt{viironen07},  \citealt{kewley01},  \citealt{mcgaugh91}, \citealt{dors08}, \citealt{dopita86}),
a comparison of oxygen estimates obtained from these diagrams and $T_{\rm e}$-method
is unavailable in the literature.

  Another important issue related with the determination of metallicity using  strong-line methods is 
the relation between the ionization parameter and the 
metallicity, which is still controversial. For example, \citet{garnett97}, in a study of the 
interstellar abundance gradient in NGC\,2403 found  
that any correlation between ionization parameter and abundance must be a weak one. 
This result is in agreement with that obtained by \citet{kennicutt96}
from measurements of the [S II]/[S III] ratio in 41 H II regions in M101 and with the one found
for three barred galaxies established by \citet{dors05}. On the other hand,  \citet{bresolin99} found
that in metal poor disk \ion{H}{ii} regions  the ionization parameter is about
4 times larger than in \ion{H}{ii} regions with solar metallicity.
This relation  was also found by \citet{maier06} and \citet{nagao06}  for a sample of galaxies.  
This subject is important  to study the association of the 
mass-metallicity relation with the mass-age relation in local galaxies
or  the relation between gas metallicity and stellar metallicity
\citep{nagao06}. Additional analysis using several methods
would help to solve this disagreement.

\begin{figure}
\centering
\includegraphics[angle=-90,width=9cm]{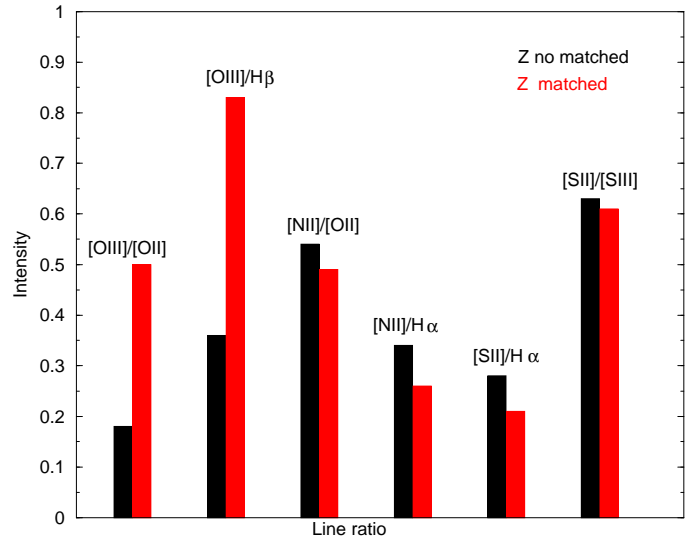}
\caption{Comparison of some predicted line ratio intensities from photoionization models
with nebular and stellar metallicity matched (red) and no matched (black). }
\label{f1a}
\end{figure}

In this paper, we employed a grid of photoionization models  and
data compiled  from the literature in order 
to estimate the oxygen abundance   
using   diagnostic diagrams   to compare with 
those via $T_{\rm e}$-method, as well as to investigate the relation
of the ionization parameter with the metallicity. Detailed photoionization
models are also built to produce more precise determinations of these parameters.  
In Section ~\ref{obs}, we describe the observational data used in the analysis.
The modeling procedures are presented in  Sect.~\ref{phot}. A description
of the diagnostic diagrams employed is given in Sect.~\ref{diag}.
The results and discussion are presented in Sects.~\ref{res} and
\ref{disc}, respectively. A conclusion of the outcome is
given in Sect.~\ref{conc}.

\section{Observational data}
\label{obs}

 Observational emission line intensities of 
a sample of \ion{H}{ii} galaxies and \ion{H}{ii} regions  and the oxygen abundance
computed using  the $T_{\rm e}$-method were compiled from literature.

The emission lines considered in our analysis
are listed in Table~\ref{tab1}.
This compilation includes  data of \ion{H}{ii} regions obtained by \citet{bresolin09}, \citet{bresolin05}, \citet{bresolin04}, \citet{bresolin07}, \citet{lee04}, and \citet{kennicutt03}. 
The data on \ion{H}{ii} galaxies were
obtained by \citet{izotov06}, \citet{leej04}, \citet{izotov04}, \citet{vilchez03}, \citet{hagele}, and \citet{guseva00}.
The sample consists of   446 objects (86 \ion{H}{ii} regions and 360 \ion{H}{ii} galaxies)
whose O/H abundance are in the range $\rm 7.0 < 12+log(O/H)_{\it{T}_{\rm e}} < 9.0$ and represents 
practically the entire  metallicity range of star-forming regions (see \citealt{pyliugin04a}).
The objects of the sample have $z < 0.07$ and   measurements
corrected by dust extinction and no AGN and gas shock contributions are present in their
ionization.

\begin{table}
\caption{Emission line ratios considered}
\label{tab1}
\begin{tabular}{ll}
\noalign{\smallskip}
\hline 
\noalign{\smallskip}
Symbol    &  Definition \\
\noalign{\smallskip}
\hline 
\noalign{\smallskip}
$R_{23}$    & ([\ion{O}{ii}]$\lambda$3727+[\ion{O}{iii}]$\lambda$4959,$\lambda$5007)/H$\beta$\\
$[\rm O\,III]$/H$\beta$ & $[\rm O\,III]$$\lambda$5007/H$\beta$\\
$[\rm O\,III]$/$[\rm O\,II]$   & $[\rm O\,III]$$\lambda$5007/$[\rm O\,II]$$\lambda$3727 \\
$[\rm N\,II]$/H$\alpha$ & $[\rm N\,II]$$\lambda$6584/H$\alpha$\\
$[\rm N\,II]$/$[\rm O\,II]$ & $[\rm N\,II]$$\lambda$6584/$[\rm O\,II]$$\lambda$3727\\
$[\rm S\,II]$H$\alpha$ &  ($[\rm S\,II]$$\lambda$6716+$\lambda$6731)/H$\alpha$\\
$[\rm S\,II]$/$[\rm S\,III]$ & ($[\rm S\,II]$$\lambda$6716+$\lambda$6731)/($[\rm S\,III]$ $\lambda$9069+$\lambda$9532)\\
\noalign{\smallskip}
\hline
\end{tabular}
\end{table}

\section{Photoionization models}
\label{phot}

\subsection{Model Grid}

\begin{table*}
\caption{Final parameters of the detailed photoionization models for \ion{H}{ii} regions observed by \citet{kennicutt03}}
\label{tab2}
\begin{tabular}{lcccccc}
\noalign{\smallskip}
\hline 
\noalign{\smallskip}
\ion{H}{ii} region     & 12+log(O/H) &  log(N/O) & log(S/O) & log\,$U$ & $N_{\rm e} \: \rm (cm^{-3})$ & Age\,(Myr) \\
\noalign{\smallskip}
\hline 
\noalign{\smallskip}
H\,1013    	       &      8.48	    &	-0.72    &	-1.34       &   -1.94 &  47   & 2.5  \\ 
H\,1105                &      8.71  	    &	-1.06    &	-1.64       &	-2.35 &  248  &  1.0 \\ 
H\,1159                &      8.80  	    &	-1.28    &	-1.94       &	-2.55 &  5  &  1.0 \\ 
H\,1170                &      8.07  	    &	-0.83    &	-1.31       &	-2.60 & 9   & 2.5  \\ 
H\,1176                &      8.16  	    &	-0.76    &	-1.38       &	-2.05 &  33  &  2.5 \\ 
H\,1216                &      8.00  	    &	-1.28    &	-1.64       &	-2.52 &  33  &  1.0 \\ 
H\,336                 &      8.75  	    &	-0.80    &	-1.58       & 	-2.68 &  15  &  2.5 \\ 
H\,409                 &      8.53  	    &	-1.15    &	-1.58       &	-2.37 &  213  & 1.0  \\ 
H\,67	               &      8.00  	    &	-1.15    &	-1.64       &	-2.82 &  10  & 2.5  \\ 
N\,5471-D              &      8.10  	    &	-1.15    &	-1.64       &	-2.4  &  110  &  2.5 \\ 
S\,DH323               &      7.76  	    &	-1.45    &	-1.61       &	-2.91 &  61  &  2.5 \\ 
\noalign{\smallskip}
\hline
\end{tabular}
\end{table*}

\begin{table*}
\caption{Observed intensity lines and those predicted by our detailed models}
\label{tab3}
\begin{tabular}{lcccccccccccc}
 \hline 
\noalign{\smallskip}
\ion{H}{ii} region & \multicolumn{2}{c}{$[\rm O\,II]\lambda3727$} &  \multicolumn{2}{c}{$[\rm O\,III]\lambda4363$} &
\multicolumn{2}{c}{$[\rm O\,III]\lambda5007$}& \multicolumn{2}{c}{$[\rm N\,II]\lambda6584$} & 
\multicolumn{2}{c}{$[\rm S\,II]\lambda6720$} & \multicolumn{2}{c}{$[\rm S\,III]\lambda9069+\lambda9532$} \\
\noalign{\smallskip}	   
\hline
\noalign{\smallskip}
            & Obs.  &Mod. & Obs. &Mod.  &Obs.&Mod.&Obs. &Mod. &Obs.&Mod.&Obs.&Mod.\\
\noalign{\smallskip}	   
\hline
\noalign{\smallskip}	    
H\,1013     & 188$\pm10$  &185   &--- & 0.20 &103$\pm5$  & 105  &64.6$\pm3.4$  &67   &28.8$\pm1.1$&26  &131.6$\pm7.0$ &171 \\
H\,1105     & 185$\pm10$  &188   &1.4$\pm0.1$ & 0.74&316$\pm17$ & 317  &33.4$\pm1.8$  &34   &24.1$\pm0.9$& 27 &126.6$\pm11$  &139 \\
H\,1159     & 198$\pm10$  &194   &1.9$\pm0.4$ & 0.63 &317$\pm17$ & 316  &23.6$\pm1.3$  &25   &29.8$\pm1.1$ & 27 &97.0 $\pm5.5$ &66 \\
H\,1170     & 308$\pm16$  &295   &1.6$\pm0.2$ & 1.67 &201$\pm11$ & 194  &44.0$\pm2.3$  &41   &56.7$\pm2.1$ & 51 &170.0$\pm9.3$ &136 \\
H\,1176     & 160$\pm8$   &153   &2.4$\pm0.3$ &2.74 &369$\pm20$ & 357  &21.2$\pm1.1$  &23   &23.0$\pm0.8$ & 19 &113.5$\pm6.1$ &137 \\
H\,1216     & 151$\pm8$   &149   &4.7$\pm0.3$ &6.67 &473$\pm25$ & 478  &11.0$\pm0.6$  &9    &18.9$\pm0.7$ & 18 &83.0$\pm4.6$  & 63\\
H\,336      & 178$\pm9$   &183   &--- &0.02 &23$\pm1$   & 26   &95.9$\pm5.1$  &99   &56.8$\pm2.1$ & 58 &107.0$\pm5.7$ & 102\\
H\,409      & 218$\pm12$  &212   &2.3$\pm0.2$ &1.39 &370$\pm20$ & 359  &27.3$\pm1.4$  & 24  &31.2$\pm1.1$ & 28 &90.1$\pm4.9$  &141 \\
H\,67       & 244$\pm13$  &248   & 3.5$\pm0.5$& 4.67&342$\pm18$ & 356  &16.3$\pm0.9$  & 19  &26.3$\pm1.3$ & 29  &92.1$\pm5.7$  & 62 \\
N\,5471-D   & 137$\pm7$   & 140  &8.0$\pm0.4$ &7.10 &578$\pm31$ & 574  &8.5$\pm0.5$   & 11  &20.6$\pm0.8$ & 18 &75.7$\pm4.3$  & 75\\
S\,DH323    & 194$\pm10$  & 198  & 5.5$\pm0.9$&3.7 &227$\pm12$ & 234  &7.9$\pm0.7$   &  7  &20.8$\pm1.4$ & 23 &---	     &42 \\
\noalign{\smallskip}	   
\hline
\end{tabular}
\end{table*}

\begin{figure*}
\centering
\includegraphics[angle=-90,width=\textwidth]{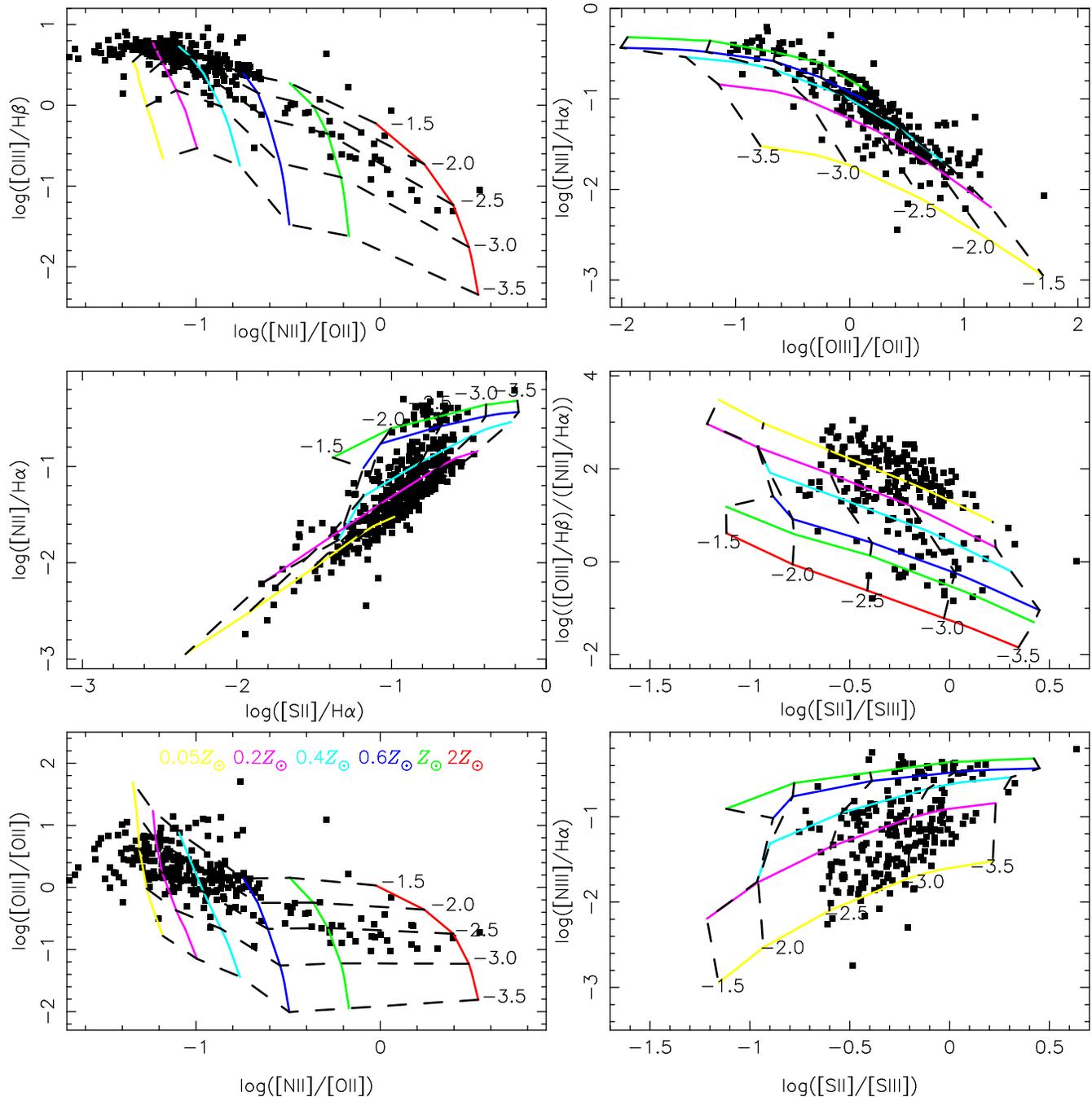}
\caption{Diagnostic diagrams containing observational data  taken from literature
(see Sect. \ref{obs}) and results of the grid of photoionization models (see Sect.\ref{phot}).
The solid lines connect curves of iso-$Z$, while
the dotted lines connect curves of iso-$U$. The values of $\log U$ and $Z$
are indicated. Squares are the observational data. The line ratio used in each plot
is defined in Table~\ref{tab1}. The typical error bar (not shown) of the emission line ratio is
about 10 \%.}
\label{f1}
\end{figure*}

To enable the estimation of the oxygen abundance and the ionization parameter using diagnostic diagrams,
we  employed the photoionization code Cloudy 8.00 \citep{ferland02}
to build a grid of models covering a large space of nebular parameters.
In these models, a stellar cluster was assumed to be responsible for the ionization of the nebulae, with
a spectral energy distribution (SED)
obtained using $Starburst99$ \citep{leitherer99}. We built models with stellar
cluster formed  by instantaneous burst with
Salpeter initial mass function ($\alpha = -2.35$), lower and upper  stellar  
mass limits of 0.1 $M_{\odot}$ and 100 $M_{\odot}$, respectively, and  age of  2.5 Myr.  
Other papers that have considered stellar clusters as ionizing sources
in order to reproduce strong forbidden lines of \ion{H}{ii} region  
  \citep{dors05, dopita00, stasinska03, bresolin99, copetti85} have derived about the
  same age for star-forming regions (i.e. 1-3 Myr). Similar ages have also been
found from optical photometric data of giant \ion{H}{ii} regions (e.g. \citealt{mayya96}). 
Selection effects may explain this limited
range of ages. \ion{H}{ii} regions younger than about 1 Myr are
difficult to be detected in the optical, because they are generally embedded
in dusty molecular clouds which cause considerable optical
extinction. Nebulae older than about 5 Myr are also difficult
to be observed because their original massive stars have cooled or
are dead  \citep{dopita00, garcia96, copetti85}.  
We used the stellar  evolution models from the Geneva group   with high mass-loss rates
 and without stellar rotation \citep{meynet94}. The non-LTE atmosphere model of \citet{pauldrach01} was assumed
 in the models.
If LTE atmosphere model is assumed instead of non-LTE model a lower ionization degree is produced in the hypothetical
nebulae \citep{dors03,stasinska97}. This would affect mainly ionization parameter  rather than metallicity determinations via
strong-line methods. 
The models were built having ionization
parameter ranging from $\log U =-1.5$ to $-3.5$ (with a bin size of 0.5 dex), metallicities (traced by the oxygen abundance) 
$Z$= 0.04, 0.02, 0.008, 0.004, 0.001, plane-parallel geometry, and 
electron density of $N_{\rm e}= 200 \: \rm cm^{-3}$. This electron density value is typical
of not evolved  \ion{H}{ii} regions  \citep{copetti00}.

The abundances of heavy metals in the nebula is scaled linearly to the solar metal
composition  through the comparison of the oxygen abundances,
with the exception of the N abundance,
which was taken from the relation log(N/O)=log(0.034+120\,O/H) of \citet{vila93}.  
The solar composition ($Z$= 0.02) refers to \citet{allende01} 
and correspond to 12+log(O/H)= 8.69. 
The presence of internal
dust was considered and the grain abundances \citep{hoof01} were also linearly scaled with the oxygen abundance.
To take depletion of refractary elements onto dust grains
into account the abundances of the elements Mg, Al, Ca, Fe,
Ni, and Na were reduced by a factor of 10, and Si by a factor
of 2 \citep{garnett95} relative to the adopted abundances in
each model.

 The solar metallicity for the stars from the Geneva evolutionary tracks, which corresponds to the old solar oxygen 
abundance value [12+log(O/H)= 8.87 \citep{grevesse98}],  is 
higher than the value adopted for the nebular component. This produces  an imperfect match
between gas and star metallicity  in our photoionization models. As pointed out by 
\citet{dopita06}, the main effect of this is that the computed stellar UV photon field result slightly softer.  
To investigate how this discrepancy affects our results, 
we built a model with a perfect match between nebular and stellar metallicities  and 
compared the result with another model whose metallicities are
in disagreement.
This latter was built using a SED with 12+log(O/H)=8.69 obtained by
linear interpolation of the spectra with $Z$= 0.02 and $Z$= 0.008. 
In Figure~\ref{f1a} we show the histogram with the
comparison of some emission line intensities ratio predicted by these models.
We can see that, with exception of  \ion[O\,III]/\ion[O\,II]
and \ion[O\,III]/H$\beta$, the intensities of the majority of the line ratios 
show little variation when the stellar metallicity atmosphere change.
  Similarly,  the predicted value of the $R_{23}$ index (not shown in Fig.~\ref{f1a})  
 shows a small deviation, i.e. about 12\%, 
which corresponds to variations in the oxygen abundance from calibrations using this line ratio
 by only 0.02 dex.
Thus, the disagrement between the nebular and stellar metallicity  have little influence on the
$Z$ determinations from strong-line methods and can only affect $U$ estimates in diagnostic
diagrams which use the  \ion[O\,III]/\ion[O\,II]
and \ion[O\,III]/H$\beta$.   
Interestingly, \citet{dopita06} found that critical line ratios changed  by 0.1 dex or less, 
except for the [\ion{O}{i}]$\lambda$6300/H$\alpha$ ratio, when a test model with a 
0.4 $Z_{\sun}$ spectral synthesis cluster model from   $Starburst99$ embedded in a nebula with 1.0 $Z_{\sun}$ (12+log(O/H)= 8.66) is run.
Along the paper the solar abundance 
adopted refers to   12+log(O/H)= 8.69 from \citet{allende01}.

These models are similar to the ones of \citet{dors06} and have been successful in describing  observational data of \ion{H}{ii} regions (see \citealt{dors08};  \citealt{krabbe08}, \citealt{krabbe07}).

\subsection{Detailed models}

In general,   grids of photoionization models are built assuming a fixed  
N/O-O/H  relation. However, this constancy can yield large uncertainties in O/H estimates via strong-line methods  \citep{perez09}. This problem can be circumvented by the use of detailed photoionization models.
To analyze the source of these uncertainties, we built detailed models in order to reproduce the observational emission line intensities of  11
\ion{H}{ii} regions (see Table~\ref{tab2}) located along the disk of the galaxy M\,101, observed
 by \citet{kennicutt03}, 
and we compared our estimates with O/H and $U$ values from other methods.   These objects were selected because 
they cover the wide  range in metallicity and ionization parameter considered in this paper.
We computed individual models for each object adopting the following methodology.
Firstly, a model for each region was built by initially
 guessing  the  $Z$ and  $U$ values
derived from a comparison between the  grid of photoionization models shown 
in the diagnostic diagram [\ion{O}{iii}]/[\ion{O}{ii}] vs. [\ion{N}{ii}]/[\ion{O}{ii}]
(see Figure~\ref{f1})  and the observational data. The electron density
of each model was considered to be that computed utilizing the task  temden
of the package IRAF, where we consider the  sulfur ratio 
[\ion{S}{ii}]$\lambda$6716/[\ion{S}{ii}]$\lambda$6731 and  electron temperature
for the $\rm O^{+}$ ion measured by \citet{kennicutt03}.
The stellar cluster was assumed to have an age of 2.5 Myr, $M_{\rm up}$ = 100 $M_{\odot}$
and metallicity was matched with the closest one nebular assumed in the models.
Then, we ran new models ranging the O/H and $U$ values by 0.3 and 0.5 dex, respectively, with a step of 0.1 dex. 
From this series of models we selected a model which produced the smallest 
$\sum \chi_{i}^2=\chi_{[\rm O\, II]/\rm H\beta}^2 +\chi_{[\rm O\,III]/\rm H\beta}^2$,
where  $\chi_{i}=(I_{\rm obs.}^{i}-I_{\rm pred.}^{i})^{2}/I_{\rm obs.}^{i}$;
$I_{\rm obs.}^{i}$ and $I_{\rm pred.}^{i}$ are the observational and predicted
intensities of the line ratios, respectively. Another series of models was computed
considering the O/H and $U$ values found by the criterion above but ranging the N/H and S/H abundances
by 0.3 dex in order to reproduce the intensities of the
[\ion{N}{ii}]$\lambda$6584 and [\ion{S}{ii}]$\lambda$6720 emission lines. 
The satisfactory solution is found when  $I_{\rm pred.}$
reproduces $I_{\rm obs.}$ within the observational uncertainties and the model 
has the smallest $\sum \chi_{i}^2=\chi_{[\rm O\,II]/\rm H\beta}^2 +\chi_{[\rm O \,III]/\rm H\beta}^2
+\chi_{[\rm N \,II]/\rm H\beta}^2 +\chi_{[\rm S\, II]/\rm H\beta}^2$.
In some cases no satisfactory solution was reached 
considering the age of the ionizing cluster of 2.5 Myr. For these, it 
was necessary to assume an age of 1 Myr because their observed emission lines could
only be reproduced by means of a harder spectral energy distribution.  
In Table~\ref{tab2}, we present
the   final parameter obtained  for the models.

\begin{figure}
\centering
\includegraphics[width=\columnwidth]{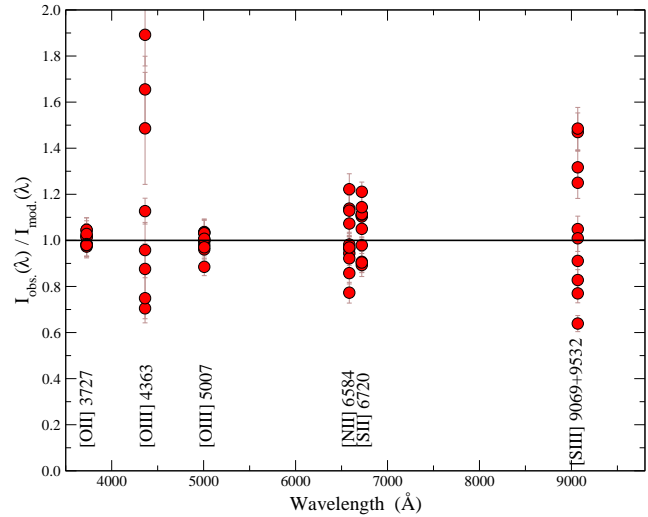}
\caption{Ratio between observed and predicted emission line intensities of some \ion{H}{ii} regions
located in M\,101 observed by \citet{kennicutt03}.}
\label{f66}
\end{figure}

\section{Diagnostic Diagrams}
\label{diag}

\begin{figure}
\centering
\includegraphics[angle=-90,width=8cm]{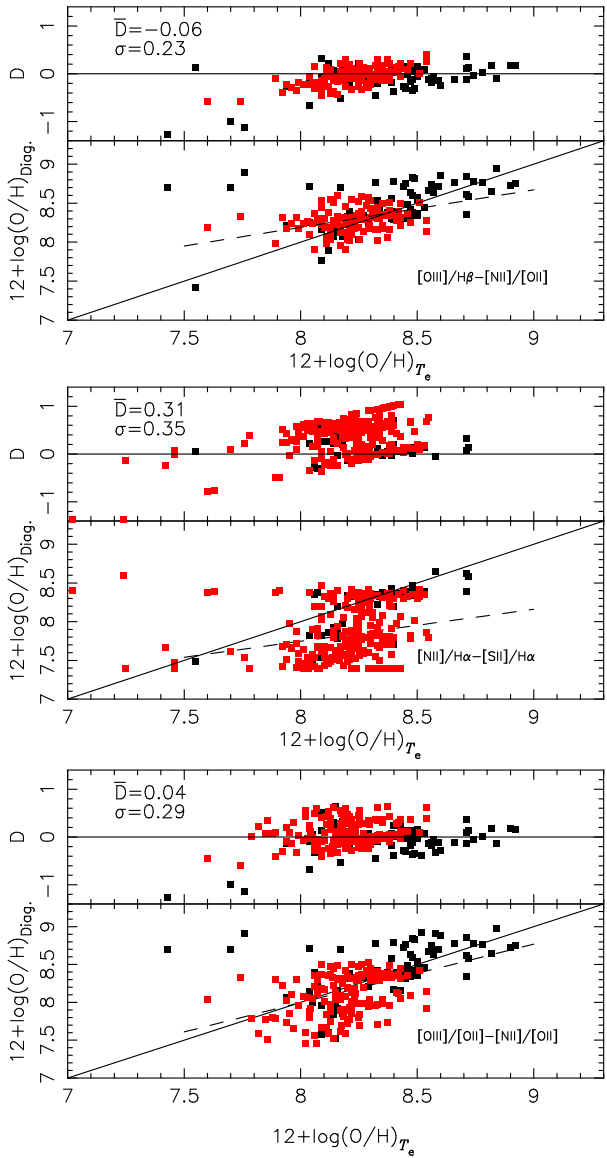}
\caption{Comparison between the oxygen abundances derived using the $T_{\rm e}$-method with
those computed using the diagnostic diagrams. In each plot  the diagnostic
diagrams used to compute the oxygen abundance are indicated. The top panel of each diagram shows the
difference between oxygen abundance via $T_{\rm e}$-method and via diagnostic diagrams.
The average value of this difference and the dispersion are shown in each plot.
Solid lines represent the equality of the two estimates. The results for \ion{H}{ii} galaxies
and \ion{H}{ii} regions are marked by  red and black  points, respectively}
\label{f3}
\end{figure}

We employ six diagnostic diagrams containing predicted and observed emission line ratios 
sensitive to  $Z$ and $U$. The diagrams considered are described below.

\begin{itemize}
 
\item $ $[\ion{O}{iii}]/[\ion{O}{ii}] vs. [\ion{N}{ii}]/[\ion{O}{ii}] --- Diagnostic diagram suggested by 
\citet{dopita00}, where the [\ion{O}{iii}]/[\ion{O}{ii}] has a strong dependence on $U$,
 on the effective temperature of the ionizing stars
(e.g. \citealt{dors03}; \citealt{perez05}) and on the metallicity \citep{kewley02, dopita00}. 
The [\ion{N}{ii}]/[\ion{O}{ii}]  correlates strongly with $Z$ above to $Z > 0.04 \:Z_{\odot}$ \citep{kewley02},
it is also dependent  on the N/O abundance ratio \citep{perez09} and it is almost independent of $U$ \citep{kewley02}.

\item $  $[\ion{N}{ii}]/H$\alpha$ vs. [\ion{S}{ii}]/H$\alpha$ --- This diagram was proposed
by \citet{viironen07}  to estimate the metallicity, where both line ratios
are dependent on $U$ and $Z$  \citep{thaisa94,kewley02}.
\citet{mazzuca06} pointed out that the use of diagnostic diagrams using [\ion{N}{ii}]/H$\alpha$ can yield degenerate values for $Z$, as star-forming regions
with low $Z$ and $U$ have [\ion{N}{ii}]/H$\alpha$ values similar to regions with 
high $Z$ and $U$.   In addition, no consistent values were found for oversolar abundances because 
the [\ion{N}{ii}]/H$\alpha$ parameter saturates in this high-metallicity regime.
Therefore,  we did not consider models with metallicity over-solar. For the [\ion{S}{ii}]/H$\alpha$, this line ratio is strongly dependent on $U$ and increases with the abundance for low metallicities \citep{levesque10}.

\item $  $ [\ion{O}{iii}]/H$\beta$ vs. [\ion{N}{ii}]/[\ion{O}{ii}] --- Diagram proposed
by \citet{dopita00}.  The [\ion{O}{iii}]/H$\beta$  
was suggested by \citet{edmunds84} as an O/H indicator. However, due to its dependence on  $U$  \citep{dopita86,mcgaugh91}
a combination with another line ratio is preferable, otherwise, crude O/H estimates with
uncertainties of about 0.5 dex are produced \citep{kobulnicky99}.

\item $  $ [\ion{N}{ii}]/H$\alpha$ vs. [\ion{S}{ii}]/[\ion{S}{iii}] ---  
The [\ion{S}{ii}]/[\ion{S}{iii}] was proposed to be a $U$ indicator
by \citet{diaz91} for  moderate to high metallicity regime 
(see also \citealt{dopita86}), and it has   little 
dependence on $Z$. The problem in using
this line ratio is that it is underestimated by photoionization models \citep{garnett89},
especially  for high metallicity \citep{dors05} doing any estimation  in this regime is somewhat uncertain.

\item $  $([\ion{O}{iii}]/H$\beta$)/([\ion{N}{ii}]/H$\alpha$) vs. [\ion{S}{ii}]/[\ion{S}{iii}] --- 
\citet{pettini04} showed that ([\ion{O}{iii}]/H$\beta$)/([\ion{N}{ii}]/H$\alpha$) is
dependent on  $Z$.
Because this line ratio is also dependent on $U$, we combined it with the [\ion{S}{ii}]/[\ion{S}{iii}]
in order to minimize the uncertainties in $Z$ determinations.

\item $  $[\ion{N}{ii}]/H$\alpha$ vs. [\ion{O}{iii}]/[\ion{O}{ii}] --- We  investigate
the combination between these line ratios to eliminate the problem existing with 
the use of  [\ion{S}{ii}]/[\ion{S}{iii}]. 
\end{itemize}

\section{Results}
\label{res}

\begin{figure*}
\centering
\includegraphics[angle=-90,width=\textwidth]{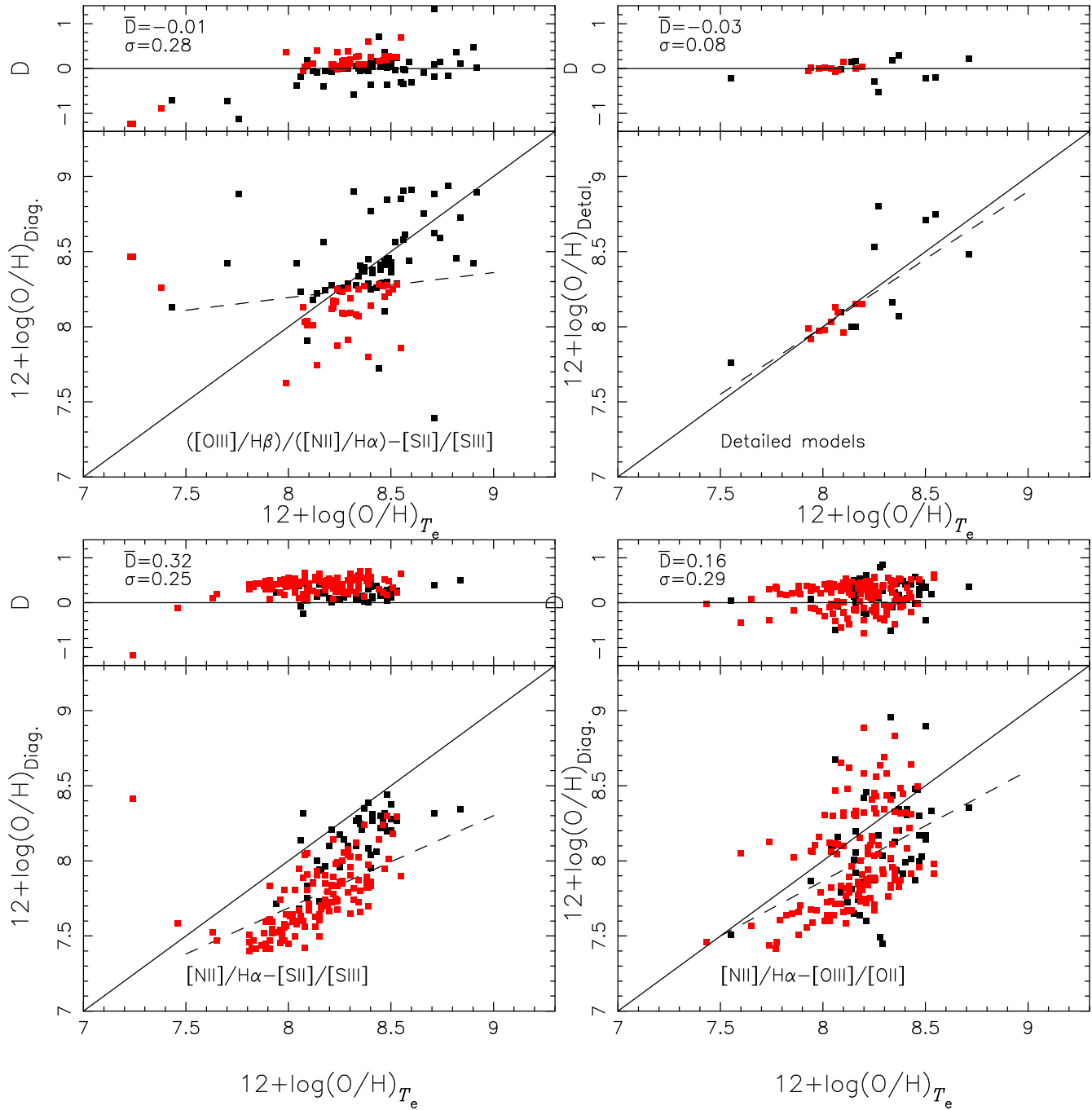}
\caption{Same as Fig.~\ref{f3}, but for other diagnostic diagrams. The panel in the right top
represents the comparison using detailed photoionization models and  the red points
were taken from \citet{perez10}.The results for \ion{H}{ii} galaxies and \ion{H}{ii} regions
are indicated by different symbols (red and black dots, respectively).}
\label{f4}
\end{figure*}

In Figure~\ref{f1}, the diagnostic diagrams described  above containing 
the results of our grid of photoionization models and the data sample is shown.
 The majority of the observational data fall
within  the regions occupied by the models. However, in the 
diagram   [\ion{O}{iii}]/H$\beta$ vs. [\ion{N}{ii}]/[\ion{O}{ii}]    
the models predict  [\ion{O}{iii}]/H$\beta$ values lower than the observed ones
for the low metallicity regime  and  high $U$ values,   result also found for the
([\ion{O}{iii}]/H$\beta$)/([\ion{N}{ii}]/H$\alpha$)  line ratio. 
Similar difficulty in modeling metal poor star-forming regions  have been found
by other authors. For example,  \citet{dopita06}, who used the same  SEDs used in this paper, 
found  that their models do not reproduce the observed emission line diagnostic ratios
of objects with $Z < 0.4 \: Z_{\odot}$. \citet{martinmajon08}, using a combination of photoionization
models  and sets
of stellar yields from \citet{gavilan05}, also found that their models  
do not reproduce observational data  of most metal deficient \ion{H}{ii} galaxies
 (see also \citealt{fernandes03}).
\citet{kewley01} pointed out that this disagreement is due to that  stellar ionizing spectra
are not hard enough in the far ultraviolet region, and  inclusion of the effects of continuum
metal opacities in stellar atmospheres should be a way of improving the models accuracy. However, in our work 
we adopted  stellar atmosphere models  of \citet{pauldrach01}, which include treatments of continuum
metal opacities, and still this  disagreement  can be noted.  
A solution for this problem seems to include the effects of rotation
in stellar models (see discussion above). 
The number of points ranges in the diagrams because for some data set no all
emission lines considered were observed. For example, in \citet{izotov06} the
[\ion{O}{ii}]$\lambda$3727 is not observed in 30\% of the objects.

 For the detailed models,   a comparison of the  
predicted and observed emission line intensities is listed in Table~\ref{tab3},  and
  Fig.~\ref{f66} shows the ratio between  these.
We can see that the
models reproduce very well (with differences lower than $\sim 15$ \%) all the
observed intensities within of the observational uncertainties, with exception
of the [\ion{S}{iii}]$\lambda9069+\lambda9532$  and [\ion{O}{iii}]$\lambda4363$ emission lines,
which are reproduced only for H\,336 and NGC\,5471-D; and for NGC\,1170 and NGC\,1176, respectively.
  Other works have also found that photoionization models 
 are   incapable of reproduce  
 emission line intensities sensitive to $T_{\rm e}$  (e.g. \citealt{stasinska99, oey00}).
  This problem has been atributted to temperature gradients  and/or
 temperature inhomogeneities in nebulae which are not taken into account in simple
 photoionization models (see \citealt{stasinska02}) such as the ones used in this paper.
 Because the [\ion{O}{iii}]$\lambda4363$ emission line has an
 exponential dependence on the electron temperature, any  offset between the
 electron temperatures in  photoionization
models and  observed forbidden-line temperatures
 will have the strong  effect on the reproducion of this emission line.

\subsection{Abundance determination comparison}
\label{abund}

To check the reliability  of diagnostic diagrams, in Figs.~\ref{f3} and \ref{f4} we present
a comparison of O/H obtained from these methods with O/H abundances 
via $T_{\rm e}$-method    
and the difference among these estimations.   We include in Fig.~\ref{f4}
the results of the detailed models for the \ion{H}{ii} regions in M\,101 as well as
the results of \citet{perez10}, who built 
detailed photoionization models in order to reproduce emission line intensities 
of 10 \ion{H}{ii} galaxies.
In each plot, the average value ($\rm \overline{D}$) and the dispersion
($\sigma$) of this difference are also presented. The O/H and $U$ values from the diagnostic diagrams
were obtained by linear interpolation from the model grid shown in Fig.~\ref{f1}. 
In a few cases double 
values of $Z$ and/or $U$  for a same objects are found
because the models overlap for a given combination of these parameter. For these,
the estimations  were not considered in our analysis. This happened    
mainly for the [\ion{N}{ii}]/H$\alpha$ vs. [\ion{S}{ii}]/H$\alpha$ diagnostic diagram
(for about 5\% of the points).

\begin{figure*}
\centering
\includegraphics*[angle=-90,width=\textwidth]{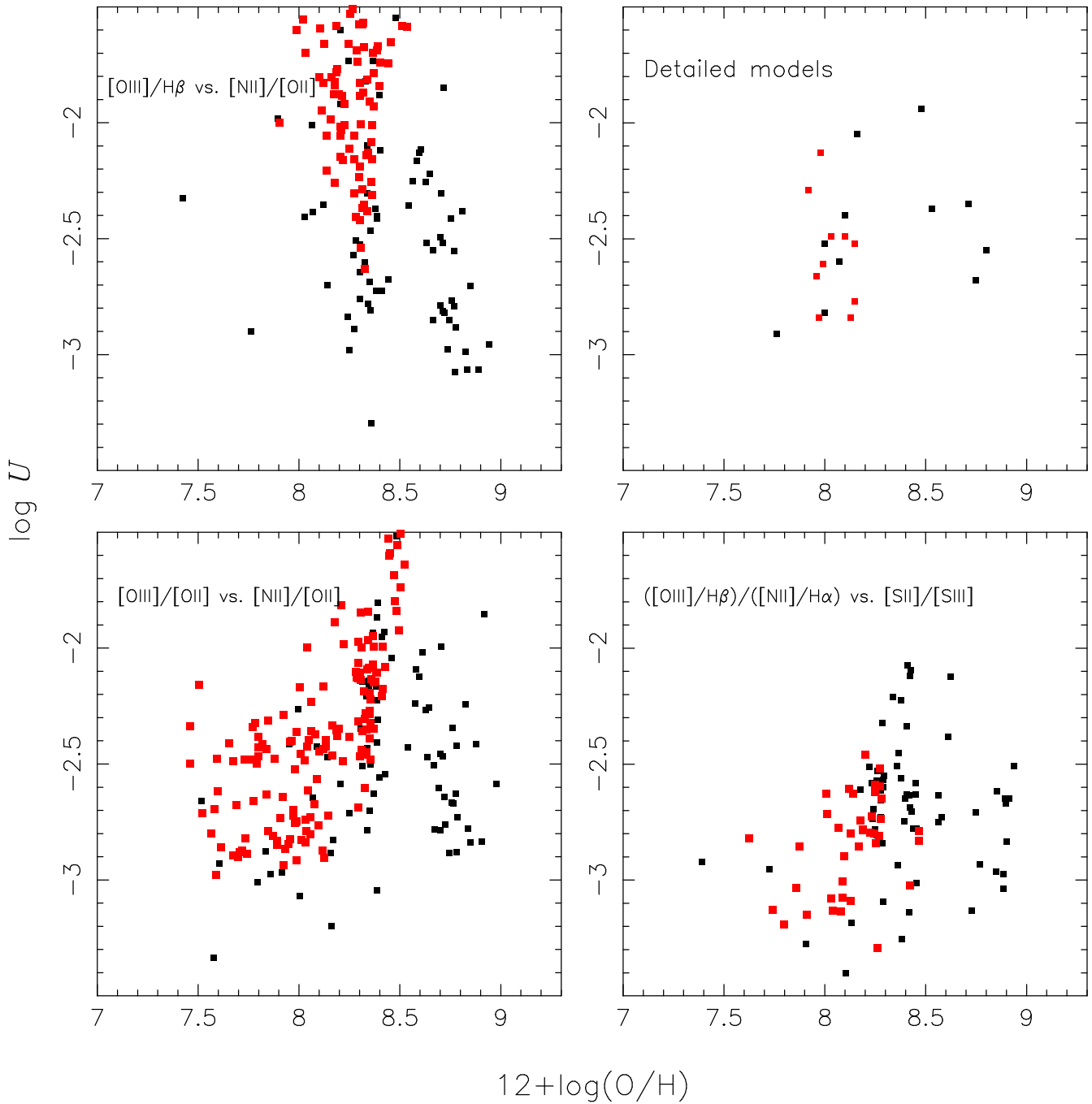}
\caption{Logarithm of ionization parameter vs. oxygen abundances. Points
represent  the results using the diagnostic diagrams indicated in each plot.
The results for \ion{H}{ii} galaxies
and \ion{H}{ii} regions are marked by symbols with different colors (red for \ion{H}{ii} galaxies
and black for \ion{H}{ii} regions).
In the upper right plot the points represent estimations 
via detailed photoionization models, in which the red points were taken from \citet{perez10}.}
\label{f5}
\end{figure*}
The diagnostic diagrams that
 provide the best results are the [\ion{O}{iii}]/[\ion{O}{ii}] vs. [\ion{N}{ii}]/[\ion{O}{ii}], 
[\ion{O}{iii}]/H$\beta$ vs. [\ion{N}{ii}]/[\ion{O}{ii}], and ([\ion{O}{iii}]/H$\beta$)/([\ion{N}{ii}]/H$\alpha$) vs. [\ion{S}{ii}]/[\ion{S}{iii}], which give O/H estimates close to the  $T_{\rm e}$-method
with an absolute difference of about 0.04 dex. The  lowest dispersion is found  with the use of the
[\ion{O}{iii}]/H$\beta$ vs. [\ion{N}{ii}]/[\ion{O}{ii}] diagram. For the majority  of the diagrams,  the difference and
the dispersion are  larger in the regime of low  metallicity (12+log(O/H)$< 8.0$).  
For the other diagrams this difference is  about 0.25 dex.
The O/H abundances via detailed models are in
consonance with the ones via $T_{\rm e}$-method for the objects analyzed and the dispersion derived
is lower than the one obtained  using diagnostic diagrams.

\begin{figure}
\centering
\includegraphics[angle=-90,width=\columnwidth]{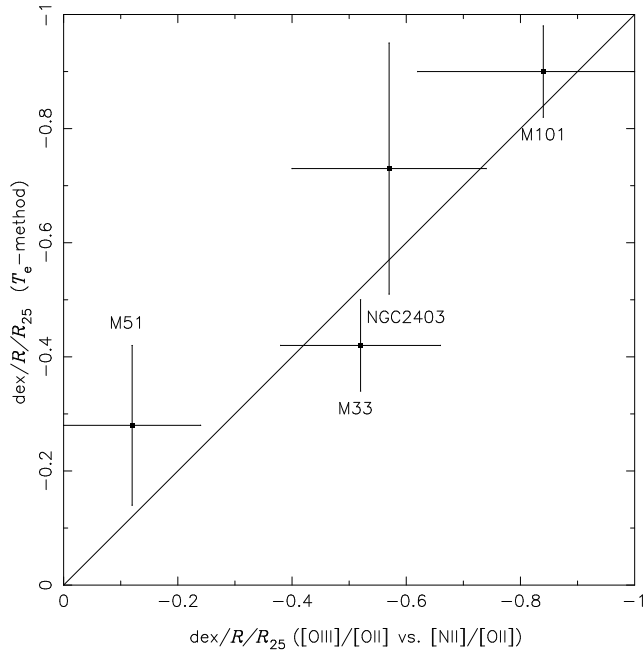}
\caption{Comparison of abundance gradient slopes derived using the [\ion{O}{iii}]/[\ion{O}{ii}] vs. [\ion{N}{ii}]/[\ion{O}{ii}] and  $T_{\rm e}$-method for some galaxies as indicated.
The slopes are measured in 12+log(O/H)/$R/R_{25}$, where  $R_{25}$  is the isophotal radius.
The authors which we collected the gradients via $T_{\rm e}$-method are cited in the text.
Solid lines represent the equality of the two estimates. }
\label{f55}
\end{figure}

\subsection{Ionization parameter determination}
\label{ion}
 
For the ionization parameter, in Fig.~\ref{f5}, we plotted $U$ against the
oxygen abundance obtained from diagnostic diagrams presented in Section~\ref{diag}
as well as those obtained from detailed models. 
The results for \ion{H}{ii} galaxies and \ion{H}{ii} regions
are indicated by different symbols (red and black dots, respectively). 
 The [\ion{O}{iii}]/H$\beta$ vs. [\ion{N}{ii}]/[\ion{O}{ii}]
and [\ion{O}{iii}]/[\ion{O}{ii}] vs. [\ion{N}{ii}]/[\ion{O}{ii}]
estimates larger $U$ values than  the ones via other methods.
There is not a clear trend of $U$ with O/H  and this result
is also confirmed  by the detailed model estimation.

\begin{figure}
\centering
\includegraphics[angle=-90,width=\columnwidth]{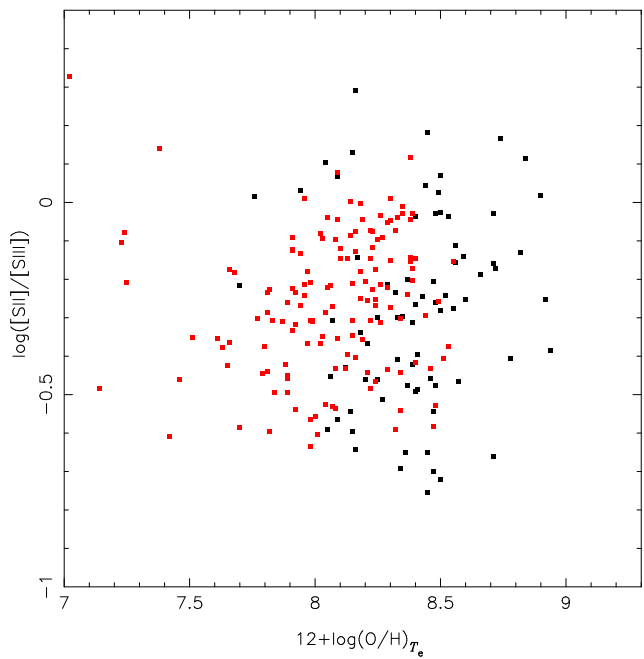}
\caption{ [\ion{S}{ii}]/[\ion{S}{iii}] line ratio observed versus oxygen abundance via $T_{\rm e}$-method
for our sample of objects. \ion{H}{ii} galaxies  
and  \ion{H}{ii} regions  are marked by  red and  black points, respectively.}
\label{f5a}
\end{figure} 

\section{Discussion}
\label{disc}

 Comparing oxygen abundance determinations  via $T_{e}$-method
for the sample of \ion{H}{ii} regions and \ion{H}{ii} galaxies with those  based
on strong emission lines, we found that   
 the [\ion{O}{iii}]/[\ion{O}{ii}] vs. [\ion{N}{ii}]/[\ion{O}{ii}], 
[\ion{O}{iii}]/H$\beta$ vs. [\ion{N}{ii}]/[\ion{O}{ii}], and ([\ion{O}{iii}]/H$\beta$)/([\ion{N}{ii}]/H$\alpha$) vs. [\ion{S}{ii}]/[\ion{S}{iii}] diagnostic diagrams  give O/H values nearest to the $T_{\rm e}$-method,
with  differences of about 0.04 dex and a dispersion of about 0.30 dex. 
This  difference    
 is about the same as the one  between oxygen estimates via
 the $P$-method \citep{pilyugin01} and via $T_{\rm e}$-method found by \citet{lopez10}
 considering a sample of Wolf-Rayet galaxies. 
 It is lower by about 0.15 dex  than the one found by these authors when using  only one emission line 
ratio sensitive to metallicity. 

As seen in Figs.~\ref{f3} and \ref{f4}, large difference are found for $\rm 12+\log(O/H)< 8.0 $
($Z\lesssim 0.2 \:Z_\odot$).
Similar results were also found by \citet{yin07}, who compared oxygen estimates via $N_{2}$
and ([\ion{O}{iii}]/H$\beta$)/([\ion{N}{ii}]/H$\alpha$) with those via $T_{e}$-method 
for a sample of 695 galaxies and \ion{H}{ii} regions.
This  occurs because in this regime of metallicity the nitrogen and oxygen have both 
mainly a primary nucleosynthesis origin,
  doing  nitrogen emission lines to be relatively independent on oxygen abundance and 
consequently  the use of metallicity indicators based on these emission lines are not reliable.
(e.g. \citealt{levesque10}; \citealt{dopita00}).

The origin of the dispersion found by us is  probable  due to the difference between the 
real N/O-O/H abundance relation  of the object sample  and the  one assumed in our models.
In fact, \citet{perez09} analyzed the dependence of N/O with  O/H estimation obtained 
via  the  metallicity indicators  using nitrogen line ratios and compared these estimations with the ones obtained  via $T_{\rm e}$-method.
They found approximately the same dispersion  as the one derived by us,
and  also showed that if the N/O ratio is taken into account in  strong-line methods, the dispersion  can be  
reduced by about 0.1 dex. Moreover, the scattering of N/O for a fixed O/H value is 
larger for the low metallicity regime 
(see e.g. \citealt{pilyugin03}), which  introduces  a larger dispersion
for oxygen estimations in  this regime, such as the one observed in our
results. This is confirmed by  the use of detailed models, for which the N/O-O/H relation
is a free parameter, yielding a lower  dispersion (0.08 dex) than the ones obtained from diagnostic diagrams.
\citet{yin07} also obtained similar results comparing oxygen abundances derived
from $T_{\rm e}$-method and those via the photoionization models of \citet{charlot01}.

Another important test is to verify if abundance gradients estimates  by using   
diagnostic diagrams agree with those via $T_{\rm e}$-method. For that, in Fig.~\ref{f55} we
show a comparison of  oxygen gradient slope computed using the [\ion{O}{iii}]/[\ion{O}{ii}] vs. [\ion{N}{ii}]/[\ion{O}{ii}]
diagram presented in Fig.~\ref{f1}
and those via $T_{\rm e}$-method for spiral galaxies M\,101, M\,51, M\,33, and NGC\,2403 obtained
 by \citet{kennicutt03}, \citet{bresolin04}, \citet{magrini07}, and \citet{garnett97}, respectively.
 We can see that, within the uncertainties given by the linear fitting, the diagnostic diagram
 above yields abundance gradient consistent with the ones via $T_{\rm e}$-method.
 Again, the difference between the gradient estimates is probable  due to the N/O-O/H  relation assumed in our models
  and the one of the galaxies. This is supported by the detailed model results, since 
 a linear fitting on oxygen abundance from these, presented in Table~\ref{tab2},  
  yields  a gradient for M\,101 of   12+log(O/H)= 0.90($\pm0.26$) $R/R_{25}$ + 8.77($\pm0.15$), the
  same gradients found by  \citet{kennicutt03} using the $T_{\rm e}$-method.

In general, oxygen determination obtained from  strong-line methods, which use emission line
intensities predicted by photoionization models, are 
overestimated  up to 0.5 dex when compared with those obtained
from $T_{\rm e}$-method \citep{kewley08,dors05,kennicutt03,garnett04,stasinska02}. 
This discrepancy is attributed  to the fact that
photoionization codes are not realistic enough, do not treat all the 
relevant physical processes correctly, use inaccurate atomic data, etc \citep{kennicutt03}.
However, as seen previously, using the state of art of photoionization models and the combination
of two line ratio, one sensitive to the metallicity and another
sensitive to the ionization parameter, which does taken into account 
the physical conditions (hardness of the ionizing radiation and geometrical factor)
of  star-forming regions \citep{pilyugin01},
minimizes the effects mentioned above and gives O/H estimates close to the $T_{\rm e}$-method.
  As explained in Sect.~\ref{phot}, the match between solar abundances for the gas and star  
has  little influence on  metallicity indicators (i.e. [\ion{N}{ii}]/[\ion{O}{ii}]) showing that
the metallicity estimates by our models are indepent of this fact.

\begin{figure}
\centering
\includegraphics[angle=-90,width=8cm]{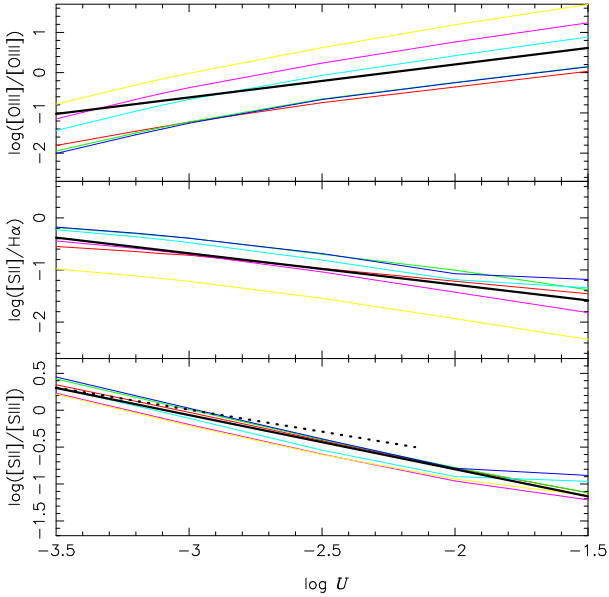}
\caption{Relation between the ionization parameter $U$ and the 
[\ion{S}{ii}]/[\ion{S}{iii}],[\ion{S}{ii}]/H$\alpha$ and [\ion{O}{iii}]/[\ion{S}{ii}]
line ratios. The colorful solid lines represent
results for different metallicities, such as in Fig.~\ref{f1}, and the black solid line represents the
 linear fitting of the average of these results. The dotted line represents the relation
proposed by \citet{diaz91}.}
\label{f6}
\end{figure}

 The ionization parameter is expected  to be  dependent on the metallicity because
stellar atmospheres of massive O stars become cooler
 with increasing metallicity   as a result of 
 enhanced line and wind blanketing
\citep{massey05}, decreasing  consequently  the ionization parameter.
Moreover, when stellar atmosphere abundance is higher, this  scatters the photons
emitted from the photosphere more efficiently, causing a greater conversion efficiency
from luminous energy flux to mechanical energy flux in the stellar wind base region,
which also leads to a diminution of $U$ in the \ion{H}{ii} region \citep{dopita06}.
A decrease of $U$ with the increase of $Z$  was found for \ion{H}{ii} galaxies
by  \citet{nagao06}   and  \citet{maier06} 
and for disk  \ion{H}{ii} regions by  \citet{bresolin99}.
However, our results indicate
no systematic dependency  of the ionization parameter
on the metallicity for the sample of objects considered, although using
our grid of models it is impossible to check what factor (e.g. stellar
effective temperature, geometrical factors, aging) is responsible for this
behavior. This result is in consonance
with the ones found by \citet{dors05}, \citet{garnett97}, and \citet{kennicutt96}, who used the  
[\ion{S}{ii}]/[\ion{S}{iii}] in order to estimate the ionization parameter in 
\ion{H}{ii} regions located in spiral disks.
To analyze whether our result is not an artefact of the methods
used in this paper, in Fig.~\ref{f5a} we plotted the [\ion{S}{ii}]/[\ion{S}{iii}]
line ratio intensity and the O/H abundances obtained by 
$T_{\rm e}$-method  of our sample. Again,
we can note no systematic behavior of $U$ with O/H.


 As noted in Fig.~\ref{f5}, the [\ion{O}{iii}]/H$\beta$ vs. [\ion{N}{ii}]/[\ion{O}{ii}]
 and [\ion{O}{iii}]/[\ion{O}{ii}] vs. [\ion{N}{ii}]/[\ion{O}{ii}] indicate
 very high $U$ values ($\log U > -2$) for some objects, which are not predicted
 by the detailed model  and by the
 ([\ion{O}{iii}]/H$\beta$)/([\ion{N}{ii}]/H$\alpha$) vs. [\ion{S}{ii}]/[\ion{S}{iii}]
 diagram. This occurs because photoionization models underpredict the [\ion{O}{iii}]/H$\beta$   
 (see Fig.~\ref{f1}), mainly for objects with  low metallicity.
 Because these  line ratios are age dependent, it is possible that models
 assuming a harder spectral energy distribution would resolve this problem. Thus,
 we  ran a grid of photoionization models (not shown) using as ionizing source a cluster of 1 Myr 
and computed  $U$ and $Z$ values. We found that, for the diagram with   [\ion{O}{iii}]/[\ion{O}{ii}], this new grid yields 
values of $U$ in consonance with the ones from other diagrams. However,   for the diagram 
with [\ion{O}{iii}]/H$\beta$, although a
better match between the models and the observational data 
was obtained,  the $U$ estimations  
continue to be overestimated in relation to the ones via other methods.
The O/H estimates practically did not change in these cases.
A simi\-lar problem was also pointed
out by \citet{stasinska03}, who using a sequence of photoionization models to reproduce observational data of  
\ion{H}{ii} galaxies, which found that the models underpredict [\ion{O}{iii}]/H$\beta$
for objects with $\rm 7.4 < \log (O/H) < 8.0$. 
These authors invoked several mechanisms to explain
this discrepancy, such as secondary ionization by X-rays and shocks but a 
definitive conclusion was not reached. Recently, \citet{levesque10}  compared 
observational data of a large sample of star-forming galaxies with
grid of photoionization models, such as the ones presented in this paper, but 
considering as ionizing source stellar cluster
with different ages and formed by instantaneous and continuous star formation.
Similar to our results, they found that very young models with 0-1 Myr and formed instantaneously	
reproduce better the observed [\ion{O}{iii}]/H$\beta$   than the older ones, although
better agreement is given by  models adopting a continuous star formation history (see also
\citealt{perez10}).  
These authors did not compare $U$ estimates obtained by different emission line ratio,
but it is probable that the discrepancy found by us is maintained even using their models
with continuous star formation.   \citet{levesque10} showed that the
 new generation of Geneva evolutionary tracks, which include  stellar rotation, produces
 a SED more prominent in the higher-energy regime ($\lambda \lesssim 230$\,\AA) than the one used here,
 with rotation  effects  being more important at lower metallicity. Thus, it is probable that
 if these SEDs were used in our photoionization models, the predicted
 intensities of the [\ion{O}{iii}]/H$\beta$ would be larger  
 and  $U$ estimates from diagnostic diagrams using this line ratio would conciliate 
with the ones from other diagrams.

As the [\ion{S}{ii}]/[\ion{S}{iii}] ratio  is weakly dependent  on $Z$, it is useful to calibrate it with $U$.
In Fig.~\ref{f6}, we show the $U$-[\ion{S}{ii}]/[\ion{S}{iii}] relation predicted by our models 
for the entire range of $Z$ and the relation proposed by \citet{diaz91},
as well as our results for [\ion{S}{ii}]/H$\alpha$ and [\ion{O}{iii}]/[\ion{O}{ii}] versus $U$.

A linear fitting of the average of these photoionization model results produce
\begin{equation}
\log U=-1.36 \:(\pm0.07) \log [\rm S\, II]/[\rm S\,III]-3.09\:(\pm 0.05),
\end{equation}

\begin{equation}
\log U=-1.66\:(\pm0.06) \log [\rm S\, II]/ \rm H\alpha-4.13\:(\pm 0.07),
\end{equation}

\begin{equation}
\log U=1.22\:(\pm0.07) \log [\rm O\, II]/[\rm O\, III]-2.25\:(\pm 0.05).
\end{equation}

As can be seen in Fig.~\ref{f6}, the [\ion{S}{ii}]/H$\alpha$ is a good $U$ indicator
because it shows little variation 
with the metallicity for $Z > 0.2 \:Z_{\odot}$ and uses  emission line  
with near wavelength, being almost independent of the reddening.  
Our result for $U$-[\ion{S}{ii}]/[\ion{S}{iii}] relation is in very good agreement 
with the relation proposed by \citet{diaz91}.

\section{Conclusion}
\label{conc}
We compared oxygen estimates for a large sample of objects 
obtained by direct detection of the electron temperature with those
via diagnostic diagrams containing strong emission lines predicted
by photoionization models, as well as  from detailed models. 
Among the diagnostic diagrams considered, we found that the ones
utilizing the emission lines [\ion{O}{iii}]/[\ion{O}{ii}] vs. [\ion{N}{ii}]/[\ion{O}{ii}], 
[\ion{O}{iii}]/H$\beta$ vs. [\ion{N}{ii}]/[\ion{O}{ii}], and [\ion{O}{iii}]/H$\beta$/[\ion{N}{ii}]/H$\alpha$ vs. [\ion{S}{ii}]/[\ion{S}{iii}] diagnostic diagrams  give O/H values nearest to the $T_{\rm e}$-method,
with  differences of about 0.04 dex and dispersion of 0.3 dex. Similar results were obtained
using detailed models  but with a smaller dispersion, of 0.08 dex. The origin of the dispersion is probably due
to differences between the real N/O-O relation of the sample and the one assumed in the
models. We did not find any correlation of the ionization parameter with the metallicity for
the objects in our sample.
We conclude that  the combination
of two line ratio predicted by photoionization models, one sensitive to the metallicity and another
sensitive to the ionization parameter, which does taken into account 
the physical conditions  of  star-forming regions,  
gives O/H estimates close to the values derived using direct detections of electron temperatures.

\section*{Acknowledgements}
This work was supported by FAPESP under  grant 2009/14787-7.
GH is grateful to the Spanish \emph{Ministerio de Educaci\'on y Ciencia} for support under grant AYA2007-67965-C03-03 and AYA2010-21887-C04-03, and the Comunidad de Madrid under grant S2009/ESP-1496 (ASTROMADRID).
EPM is grateful to the Spanish \emph{Ministerio de Ciencia e Innovaci\'on} for support under grant AYA2007-67965-C03-02 and AYA2010-21887-C04-02, and the Junta de Andaluc\'\i a under grant TIC114.
 
{}

\end{document}